\documentclass[prl,twocolumn,showpacs]{revtex4}
\usepackage{amsmath,amssymb,mathrsfs}
\usepackage{psfrag}
\usepackage{graphicx}
\usepackage{graphics}
\usepackage{epsfig}
\usepackage{bm}
\usepackage{color}
\usepackage{verbatim,color,ulem}

\newcommand{\beq}{\begin{equation}}
\newcommand{\eeq}{\end{equation}}
\newcommand{\beqa}{\begin{eqnarray}}
\newcommand{\eeqa}{\end{eqnarray}}

\begin{document}

\title{Non-Universal Equation of State of the Two-Dimensional Bose Gas}
\author{L. Salasnich} 
\affiliation{Dipartimento di Fisica e Astronomia ``Galileo Galilei'', 
Universit\`a di Padova, Via Marzolo 8, 35131 Padova, Italy
\\
Consorzio Nazionale Interuniversitario per le Scienze Fisiche 
della Materia (CNISM), Unit\`a di Padova, Via Marzolo 8, 35131 Padova, Italy
\\
Istituto Nazionale di Ottica (INO) del Consiglio Nazionale 
delle Ricerche (CNR), Via Nello Carrara 1, 50019 Sesto Fiorentino, Italy}

\date{\today}

\begin{abstract}
For a dilute two-dimensional Bose gas the universal equation 
of state has a logarithmic dependence 
on the s-wave scattering length. Here we derive non-universal 
corrections to this equation of state taking account finite-range effects 
of the inter-atomic potential. Our beyond-mean-field analytical results 
are obtained performing dimensional regularization of divergent zero-point 
quantum fluctuations within the finite-temperature formalism of
functional integration. In particular, we find that in the grand canonical 
ensemble the pressure has a nonpolynomial dependence on the finite-
range parameter and it is a highly nontrivial function of chemical 
potential and temperature. 
\end{abstract}


\maketitle

\noindent
{\it Introduction}. 
The equation of state of a uniform weakly-interacting Bose gas 
has a long history. Universal beyond-mean-field theoretical results, 
which depend only on the s-wave scattering length $a_s$ of the inter-atomic 
potential, were obtained for the three-dimensional (3D) bosonic system 
by Bogoliubov \cite{bogoliubov} and 
by Lee, Huang, and Yang \cite{ly1957,lhy1957}. In one dimension (1D), 
based on a previous investigation of the 1D Bose-Fermi mapping 
\cite{girardeau1960}, Lieb and Liniger \cite{lieb1963} obtained the exact 
equation of state of a Bose gas with contact 
repulsive interaction. In the case of two spatial dimensions (2D), 
Schick \cite{schick1971} found that the equation of state of a 
uniform 2D repulsive Bose gas contains a nontrivial logarithmic term. 
This remarkable result was improved by Popov \cite{popov} who obtained 
an equation of state which, at the leading order, reduces to 
Schick's one in the dilute limit (see also 
\cite{lozovik,hines,fisher,straley}). 
More recently, Andersen \cite{andersen2d} 
and Mora and Castin \cite{mora-castin} went one step further than Popov 
finding a next-next to leading universal equation of state 
for the two-dimensional weakly-interacting Bose gas. 
It is important to stress that, in the last years, various experiments 
with ultracold and dilute atomic gases in 3D \cite{papp2008,wild2012}
and 2D \cite{salomon2010,dalibard2011} 
have put in evidence beyond-mean-field effects on the 
equation of state of repulsive bosons. Moreover, experiments on 
1D bosons \cite{kinoshita2004,paredes2004} have shown that the Lieb-Liniger 
theory is needed to accurately describe the strong-coupling 
(i.e. low 1D density) regime. 

The universal theory of the 3D weakly-interacting Bose gas 
has been extended including corrections due to the finite range 
of the inter-atomic potential 
\cite{sala-old,braaten,andersen,roth,gao,boris,pethick,ketterle}. 
These corrections give a modified Gross-Pitaevskii equation 
\cite{gao,boris,pethick,ketterle,zinner,sgarlata} for the nonuniform 
condensate and non-universal effects for quantum fluctuations 
at zero temperature \cite{braaten,andersen}. 
For a deeper understanding of the behavior of interacting 
bosonic systems in lower dimensionality, it is extremely important 
to analyze and control non-universal effect induced by the finite 
range in the equation of state also in the case of 2D and 1D Bose gases. 
In this Letter we investigate finite-range effects 
on quantum fuctuations of a 2D Bose gas 
by using the finite-temperature functional integration \cite{schakel,atland} 
on a local effective action. We derive the finite-temperature 
beyond-mean-field (one-loop, Gaussian) equation of state of the 
bosonic system performing dimensional 
regularization \cite{thooft} of zero-point energy. The final non-universal 
analytical result, which reduces to the universal Popov equation 
of state \cite{popov} in the zero-range case, exhibits a nonpolynomial 
dependence on the finite-range parameter. 

\noindent 
{\it Effective field theory for the 2D Bose gas}. 
In the study of the two-dimensional interacting Bose gas 
we adopt the path integral formalism, 
where the atomic bosons are described by a complex  
field $\psi({\bf r},\tau )$ \cite{atland}.
The Euclidean Lagrangian density of the 
system with chemical potential $\mu$ is given by
\beqa
\mathscr{L} &=& \psi^*({\bf r},\tau) \left[ \hbar \partial_{\tau}
- \frac{\hbar^2}{2m}\nabla^2  - \mu \right] \psi({\bf r},\tau)
\nonumber 
\\
&+& {1\over 2} \int d^2{\bf r}' \ 
|\psi({\bf r}',\tau)|^2 \, V(|{\bf r}-{\bf r}'|) \, 
|\psi({\bf r},\tau)|^2 \; ,  
\label{lagrangian-initial}
\eeqa
where $V(|{\bf r}-{\bf r}'|)$ is the two-body interaction potential 
between bosons. 

Given the Fourier transform ${\tilde V}(q)$ of the 
interaction potential $V(r)$ one can expand it at the second order 
in $q$ around $q=0$ finding 
\beq
\tilde{V}(q) \simeq g_0 + g_2 \ q^2 = \tilde{V}_{p,2}(q) \; ,  
\label{pseudo-potential-q}
\eeq
where 
\beq 
g_0 =  {\tilde V}(0)= \int d^2{\bf r} \, V(r) 
\label{gg_0}
\eeq
and 
\beq 
g_2 =  {1\over 2} {\tilde V}''(0) = - {1\over 4} 
\int d^2{\bf r} \, r^2 \, V(r) \; . 
\label{gg_2}
\eeq 
Thus, within this approximation where the true interatomic potential 
${\tilde V}(q)$ is substituted by the pseudo-potential $\tilde{V}_{p,2}(q)$ 
of Eq. (\ref{pseudo-potential-q}), 
the effective local Lagrangian density becomes 
\beqa 
\mathscr{L} &=& \psi^*({\bf r},\tau) \left[ \hbar \partial_{\tau}
- \frac{\hbar^2}{2m}\nabla^2  - \mu \right] \psi({\bf r},\tau)
\nonumber 
\\
&+& {g_0\over 2} |\psi({\bf r},\tau)|^4 - {g_2\over 2} 
|\psi({\bf r},\tau)|^2 \big( \nabla^2 |\psi({\bf r},\tau)|^2 \big) \; . 
\label{lagrangian-eft}
\eeqa
The term proportional to $g_2$ gives an improvement with respect to the 
contact (zero-range) approximation usually adopted in the case of 
ultracold and dilute atoms. In the three-dimensional case, 
Gaussian (one-loop) results of Eq. (\ref{lagrangian-eft}) have been obtained 
in Refs. \cite{braaten,andersen,roth}, but only at zero temperature. 
Here we investigate the two-dimensional case, which is 
nontrivial also in the absence of finite range corrections, 
both at zero and finite temperature. 

\noindent 
{\it Partition function and grand potential}. 
The partition function ${\cal Z}$ of the 
system at temperature $T$ can then be written as \cite{atland}
\beq
{\cal Z} = \int {\cal D}[\psi,\psi^*] \
\exp{\left\{ - {S[\psi, \psi^*] \over \hbar} \right\}} \; ,
\label{papo}
\eeq
where
\beq
S[\psi, \psi^*] =
\int_0^{\hbar\beta}
d\tau \int_{L^2} d^2{\bf r} \
\mathscr{L}(\psi, \psi^*)
\label{action}
\eeq
is the Euclidean action, $L^2$ is the area of the system, and
$\beta \equiv 1/(k_B T)$ with $k_B$ being Boltzmann's constant.
The grand potential $\Omega$ of the system, which is a function
of $\mu$ and $T$, is then obtained as \cite{atland}
\beq
\Omega = -{1\over \beta} \ln{\big( {\cal Z}\big)} \; .
\eeq
All the thermodynamical properties of the system can be deduced 
from $\Omega$ but, due to the interaction, some approximated procedure 
is needed to explicitly calculate $\Omega$. 

The mean-field plus Gaussian (one-loop) approximation 
is obtained setting 
\beq 
\psi({\bf r},\tau) = \psi_0 + \eta({\bf r},\tau) 
\label{shift}
\eeq
and expanding the action $S[\psi, \psi^*]$ 
of Eq. (\ref{action}) around the uniform and constant $\psi_0$ 
up to  quadratic (Gaussian) order in $\eta({\bf r},\tau)$ 
and $\eta^*({\bf r},\tau)$. 
In this way, taking into account Eq.(\ref{lagrangian-eft})   
we find the grand potential (see also \cite{andersen,schakel,sala-review})  
\beq 
\Omega(\mu,T,\psi_0) = \Omega_0(\mu,\psi_0) + \Omega_g^{(0)}(\mu,\psi_0) + 
\Omega_{g}^{(T)}(\mu,\psi_0) \; , 
\label{maipiu}
\eeq
where 
\beq 
\Omega_0(\mu,\psi_0) = 
\left( - \mu \, \psi_0^2 + {1\over 2} g_0 \, \psi_0^4 \right) \ L^2 \;  
\label{maipiu1}
\eeq
is the mean-field contribution (assuming a real $\psi_0$), 
\beq
\Omega_{g}^{(0)}(\mu,\psi_0) = {1\over 2} \sum_{{\bf q}} E_{\bf q}(\mu,\psi_0) 
\label{maipiu2}
\eeq
is the zero-point energy of bosonic excitations 
\beqa
E_{\bf q}(\mu,\psi_0) &=&
\Big[ \bigg( {\hbar^2q^2 \over 2m} -\mu
+ \psi_0^2 (g_0 +{\tilde V}_{p,2}(q)) \bigg)^2 
\nonumber 
\\
&-& \psi_0^4 {\tilde V}_{p,2}(q)^2 \Big]^{1/2} \; .
\label{spettrob0}
\eeqa
i.e. the zero-temperature contribution of quantum Gaussian 
fluctuations, while 
\beq
\Omega_{g}^{(T)}(\mu,\psi_0) =
{1\over \beta} \sum_{{\bf q}} \ln{\left(1 - e^{-\beta E_{\bf q}(\mu,\psi_0)} 
\right)} \; 
\label{maipiu3}
\eeq
takes into account thermal Gaussian fluctuations.

\noindent 
{\it Zero-temperature results}.
Imposing the crucial saddle-point condition 
\beq 
{\partial \Omega_0(\mu,\psi_0)\over \partial \psi_0} = 0 \; , 
\label{chesara}
\eeq
we get 
\beq 
\psi_0(\mu) = \sqrt{\mu\over g_0} 
\label{mu0}
\eeq 
and the following spectrum of collective excitations 
\beq 
E_{q}(\mu) = \sqrt{{\hbar^2q^2 \over 2m}
\left( \lambda(\mu) {\hbar^2q^2 \over 2m} + 2 \mu \right)} \; ,  
\label{bogo2}
\eeq
where
\beq 
\lambda(\mu) = 1+{4m\over \hbar^2} {g_2 \over g_0} \, \mu 
\label{lambda}
\eeq 
takes into account finite range effects of the inter-atomic potential. 

By using Eq. (\ref{mu0}) the mean-field grand potential (\ref{maipiu1}) 
becomes 
\beq 
\Omega_0(\mu) = - {\mu^2\over 2 g_0} \; .  
\eeq
Instead, the one-loop grand potential reads 
\beq 
\Omega_g^{(0)}(\mu) = {1\over 2} \sum_{\bf q} E_q(\mu)  \; . 
\eeq

In the continuum limit, where $\sum_{\bf q}\to L^2\int d^2{\bf q}/(2\pi)^2$, 
$\Omega_g^{(0)}(\mu)$ is ultraviolet divergent 
with $E_{q}(\mu)$ given by Eq. (\ref{bogo2}). 
This divergence can be regularized with 
dimensional regularization, where the space dimension $D$ is analytically 
continued \cite{andersen,thooft,sala-review}. 
To this end we extend the two-dimensional integral to a generic complex 
$D=2-\varepsilon$ dimension, and then take the limit $\varepsilon \to 0$. 
In this way 
\beqa 
{\Omega_{g}^{(0)} \over L^D} &=& {1\over 2} 
\int {d^{D}{\bf q} \over (2\pi)^{D}} E_{q} 
\nonumber 
\\
&=& - {A(\mu) \over 2 \kappa^{\varepsilon}} \mu^2 \ 
\Gamma(-2+{1\over 2}\varepsilon) \; ,
\label{om-eps} 
\eeqa
where the regulator $\kappa$ is a crucial scale wavenumber 
which enters for dimensional reasons: $L^D = L^2 \kappa^{\varepsilon}$.  
In Eq. $(\ref{om-eps})$ we have defined 
$A(\mu)=m/(2\pi\hbar^2 \lambda(\mu)^{3/2})$ and 
$\Gamma(z)$ is the Euler gamma function, such that 
$\Gamma(-2+\varepsilon/2)=1/\varepsilon + O(\varepsilon^0)$ 
for $\varepsilon\to 0$. 
Notice that, the strengths $g_0$ and $g_2$ of the 2D Lagrangian 
density (\ref{lagrangian-eft}) become $g_0 \kappa^{\varepsilon}$ 
and $g_2 \kappa^{\varepsilon}$ in $D$ dimensions, but the adimensional 
parameter $\lambda(\mu)$ of Eq. (\ref{lambda}) remains unchanged. 

It follows that, to leading order in $1/\varepsilon$, 
the Gaussian grand potential in $D$ dimensions reads 
\beq 
{\Omega_{g}^{(0)}(\mu) \over L^D} = - 
{A(\mu) \over 2\varepsilon \, \kappa^{\varepsilon}} \mu^2 \; . 
\label{biro}
\eeq
This expression is still divergent. Nevertheless, comparing $\Omega_g(\mu)$ 
with $\Omega_{0}(\mu)$ in $D=2-\varepsilon$ dimensions 
we find the total zero-temperature grand potential 
\beq 
{\Omega^{(0)}(\mu) \over L^D} = {\Omega_{0}(\mu)\over L^D} + 
{\Omega_g^{(0)}(\mu) \over L^D} = 
- {\mu^2\over 2 \xi_r(\mu,\kappa,\varepsilon)}  \; , 
\label{totale}
\eeq 
where it appears the ``running constant''
\beq 
{1\over \xi_r(\mu,\kappa,\varepsilon)} = 
{1\over g_0 \ \kappa^{\varepsilon}} + 
{A(\mu) \over \varepsilon \, \kappa^{\varepsilon}} \;  
\label{rodi}
\eeq
which runs by changing $\kappa$ and  depends on the dimension $D$ through 
$\varepsilon=2-D$ \cite{schakel,thooft,sala-review}. 

To remove the divergence $1/\varepsilon$ in Eq. (\ref{rodi}) we calculate the 
derivative of $1/\xi_r(\mu,\kappa,\varepsilon)$ 
with respect to $\kappa$ finding 
\beq 
{1\over \xi_r(\mu,\kappa,\varepsilon)^2}
{d\xi_r(\mu,\kappa,\varepsilon) \over d\kappa} 
= {\varepsilon\over g_0 \ \kappa^{\varepsilon+1}} 
+ {A(\mu) \over \kappa^{\varepsilon+1}} \; . 
\label{deq}
\eeq 
Now, in the limit $\varepsilon \to 0$ (i.e. $D\to 2$) we get 
\beq 
{1\over \xi_r(\mu,\kappa,0)^2} {d\xi_r(\mu,\kappa,0) \over d\kappa} = 
{A(\mu) \over \kappa} \; . 
\eeq
This first order differential equation can be easily solved by 
separation of variables, and the result is 
\beq 
{1\over \xi_r(\mu,\kappa',0)} - {1\over \xi_r(\mu,\kappa,0)} 
= - A(\mu) \ \ln{\left({\kappa'\over \kappa}\right)} \; .  
\label{simsalabim}
\eeq
We set the Landau pole 
of Eq. (\ref{simsalabim}) at the high energy scale of the system 
$\epsilon_c$, i.e. we set  $1/\xi_r(\mu,\kappa',0)=0$ at $\kappa'$ such that  
$\hbar^2\kappa'^2/(2m)=\epsilon_c$. Then, when $\kappa$ corresponds 
to the actual energy of our system, i.e. 
$\hbar^2\kappa^2/(2m)=\mu$. It follows that, 
from Eqs. (\ref{totale}) 
with $\varepsilon\to 0$ and $A(\mu)=m/(2\pi\hbar^2 \lambda(\mu)^{3/2})$ 
we obtain 
\beq 
{\Omega^{(0)}(\mu) \over L^2} = - {m \over 8\pi\hbar^2 \lambda(\mu)^{3/2}} 
\mu^2 \ \ln{\left({\epsilon_c\over \mu } \right)} \; .  
\eeq

Thus, taking into account Eq. (\ref{lambda}) and the formula 
$P=-\Omega/L^2$ which relates the pressure $P$ to the grand 
potential $\Omega$, 
we finally get the zero-temperature beyond-mean-field pressure  
\beq 
P^{(0)}(\mu) = {m \over 8\pi\hbar^2} 
{\mu^2\over 
\left( 1+ \chi \, \mu\right)^{3/2}} 
\ \ln{\left({\epsilon_c\over \mu } \right)} \;  
\label{presunto}
\eeq
where 
\beq 
\chi =  {4m\over \hbar^2}{g_2\over g_0} 
\eeq
with $g_0$ given by Eq. (\ref{gg_0}) and $g_2$ given by Eq. (\ref{gg_2}). 
Moreover, following Mora and Castin \cite{mora-castin}, we set 
\beq
\epsilon_{c} = {4\hbar^2\over m \ a_s^2 \ e^{2\gamma+1/2}} \; \; ,   
\label{epsilon_c}
\eeq
that is the high-energy scale fixed by the 2D s-wave scattering length $a_s$, 
with $\gamma \simeq 0.5772$ is the Euler-Mascheroni constant. 
Given the inter-atomic potential ${\tilde V}(q)$, the corresponding 
2D scattering length $a_s$ is obtained calculating the s-wave phase shift 
$\delta_0(q)$ that is related to $a_s$ by the 
expression \cite{mora-castin,richard,werner-castin,stoof,stoof2,burnett} 
\beq
\cot{\left( \delta_0(q) \right)} =  {2\over \pi} 
\ln{\left( {q\over 2}a_s e^{\gamma} \right)} + O(q^2) \; . 
\label{phase-shift}
\eeq
In the case of contact interaction, where 
$\chi=0$, Eq. (\ref{presunto}) reduces to the equation 
of state derived by Popov \cite{popov} from a 2D hydrodynamic Hamiltonian with 
$\epsilon_c$ an ultraviolet cutoff, which depends on the s-wave scattering 
length $a_s$ \cite{nota1}. Moreover, using Eq. (\ref{epsilon_c}), one finds 
exactly the grand potential derived by Mora and Castin 
expanding the energy in powers of a small parameter \cite{mora-castin}. 
Instead, if $\chi \neq 0$ Eq. (\ref{presunto}) 
generalizes the zero-temperature Popov's equation of state 
giving a nonpolynomial finite-range correction. 

The relative difference of the pressure (\ref{presunto}) 
with and without the finite-range correction is given by 
$|1/(1+\chi \mu)^{3/2})-1| \simeq (3/2)|\chi \mu| = 
12 \pi n R^2/|\ln{(na_s^2)}|$, by using 
$R=2\sqrt{|g_2/g_0|}$ as characteristic range of the inter-atomic 
potential \cite{nota2} and 
$\mu=8\pi\hbar^2n/(m |\ln{(na_s^2)}|)$ as leading-order 
chemical potential in terms of the gas parameter $na_s^2$ 
with $n=\partial P^{(0)}(\mu)/\partial \mu$ the 2D number density 
\cite{schick1971,popov,andersen2d,mora-castin}. 
Choosing, for example, $na_s^2=10^{-5}$ and $nR^2=6 \cdot 10^{-2}$ 
we get a correction to the pressure of about $20\%$ due to 
finite-range effects, which is much larger than the Mora-Castin  
next-next-to-leading universal correction \cite{mora-castin}
of about $2\%$ for the same value of the gas 
parameter $na_s^2$ \cite{mora-castin}. 
This regime can be experimentally achieved with $^{87}$Rb atoms, 
where $R = 1.07 \cdot 10^{-2}$ micron 
\cite{dalibard}, using $n=524$ atoms/micron$^2$ and 
tuning the 2D scattering length via Feshbach resonance \cite{dalibard2} 
to $a_s = 1.38 \cdot 10^{-4}$ micron. 
In general, given a quite small gas parameter $na_s^2$, finite-range effects 
become relevant for larger values of the non-universal adimensional 
parameter $nR^2$. In other words, sizable non-universal effects 
without next-next-to-leading 
universal corrections can be reached experimentally 
by decreasing the scattering length $a_s$ (through
Feshbach-resonance techniques) and increasing the 2D number density $n$.

Note that, instead of using Eqs. (\ref{gg_0}) and (\ref{gg_2}) 
which immediately give the parameters $g_0$ and $g_2$ knowing the 
inter-atomic potential ${\tilde V}(q)$, one can alternatively 
establish a connection between $g_0$ and $g_2$ and 
familiar low-energy scattering quantities such as 
the s-wave scattering length $a_s$ and the s-wave effective range $r_s$ 
(which is not the characteristic range $R$ of the potential). 
In two spatial dimensions this connection is 
very cumbersome and highly nonlinear \cite{richard,werner-castin}. 

\noindent 
{\it Finite-temperature results}.
The finite-temperature one-loop contribution to the equation 
of state is obtained from Eq. (\ref{maipiu2}) with Eq. (\ref{mu0}), 
which gives the finite-temperature contribution 
\beq 
P_g^{(T)}(\mu) = 
{1\over 4\pi} \int_0^{\infty} dq \, q^2 \, {dE_q\over dq} 
{1\over e^{E_q/(k_BT)}-1} \; 
\eeq
to the total pressure, within our Gaussian scheme. 
Introducing the variable $x=\beta E_q$ we get 
\beq 
P_g^{(T)}(\mu) = -{k_BT\over 4\pi} \int_0^{\infty} dx \, 
q(x,\mu,T)^2 {1\over e^x -1} \; , 
\eeq
where $q(x)$ is given by 
\beq 
q(x,\mu,T) = \sqrt{2m \mu \over \hbar^2 \lambda(\mu)} \sqrt{-1 + \sqrt{1+ 
{\lambda(\mu) (k_BT)^2 x^2\over \mu^2}}} \; . 
\eeq
Expanding this expression at low temperature $T$ we find 
\beqa 
P_g^{(T)}(\mu) &=&  \frac{1}{4\pi}
\bigg(\frac{m}{\hbar^2}\bigg)(k_B T)^3 
\bigg[ \Gamma(3)\zeta(3) 
\nonumber 
\\
&-& \Gamma(5)\zeta(5) 
\frac{\lambda(\mu)}{4\mu^2}(k_B T)^2 \bigg]
\label{ordoab}
\eeqa
where $\Gamma(x)$ is the Euler gamma function and $\lambda(\mu)$ 
is given by Eq. (\ref{lambda}). Thus, the final grand-canonical 
equation of state $P(\mu,T)$, 
that gives the pressure as a function of both the chemical 
potential $\mu$ and the temperature. Explicitly, 
\beq 
P(\mu,T)=P^{(0)}(\mu)+P_g^{(T)}(\mu) \; , 
\eeq
where $P^{(0)}(\mu)$ is given by Eq. (\ref{presunto}) 
and $P_g^{(T)}(\mu)$ is given by Eq. (\ref{ordoab}). 
As clearly shown in Eq. (\ref{ordoab}), at finite temperature $T$ 
the role of non-universal effects (which are encoded into $\lambda(\mu)$) 
increases as ratio $k_BT/\mu$ grows. This effect is somehow expected 
since the details of the potential become more relevant 
when atoms scatter at higher energy.

\noindent
{\it Conclusions}. 
We have used finite-temperature one-loop functional integration 
to obtain the non-universal equation of state of a dilute and ultracold 
gas of bosons. We have adopted an effective field theory which 
includes a low-energy finite-range contribution of the 
inter-atomic interaction. The divergent zero-point energy 
of the system has been regularized by performing dimensional 
regularization. Our analytical results at zero and finite temperature 
are highly nontrivial generalizations of old but tricky universal 
formulas \cite{schick1971,popov,mora-castin} which depend only on 
the s-wave scattering length $a_s$. 

{\it Acknowledgments}. 
The author acknowledges for partial support the 2016 BIRD project 
"Superfluid properties of Fermi gases in optical potentials" of the 
University of Padova. The author thanks Alberto Cappellaro 
and Flavio Toigo for enlightening discussions and Henk Stoof 
for useful e-suggestions.


\begin{thebibliography}{99}
\bibitem{nota1} In the case of a 2D square-well potential
$V(r)=-V_0\ \Theta(b-r)$, where $\Theta(x)$ is the Heaviside step function,  
the s-wave scattering length $a_s$ is given by 
$a_s= b \ e^{J_0(k_0)/(k_0J_1(k_0))}$ 
with $k_0=\sqrt{mV_0b^2/\hbar^2}$ and $J_0(x)$, $J_1(x)$ the 
Bessel functions of first kind.
\end{thebibliography}
\end{document}